\documentclass[12pt,preprint]{aastex}

\usepackage{graphicx}
\usepackage{amssymb}
\usepackage{natbib}
\usepackage{amsmath}
\usepackage{url}

\def\fdg{\hbox{$\ \!\!^\circ$}}

\slugcomment{}

\shorttitle{STEREO quadrature observations of the 3D structure and driver of a global coronal wave}
\shortauthors{Kienreich et al.}

\begin{document}

\title{STEREO quadrature observations of the 3D structure and driver of a global coronal wave}

\author{I.W. Kienreich, M. Temmer, and A.M. Veronig}
\affil{Institute of Physics/IGAM, University of Graz,
    Universit\"atsplatz 5, A-8010 Graz, Austria;}
    \email{ines.kienreich@uni-graz.at, mat@igam.uni-graz.at and asv@igam.uni-graz.at}

\begin{abstract}
We present the first observations of a global coronal wave (``EIT
wave'') from the two Solar Terrestrial Relations Observatory
(STEREO) satellites in quadrature. The wave's initiation site was
at the disk center in STEREO-B and precisely on the limb in
STEREO-A. These unprecedented observations from the STEREO Extreme
Ultraviolet Imaging (EUVI) instruments enable us to gain insight
into the wave's kinematics, initiation and 3D structure.  The wave
propagates globally over the whole solar hemisphere visible to
STEREO-B with a constant velocity of $\sim263 \pm 16$~km\,s$^{-1}$.
From the two STEREO observations we derive
a height of the wave in the range of $\sim$80--100~Mm. Comparison
of the wave kinematics with the early phase of the erupting CME
structure indicates that the wave is initiated by the CME
lateral expansion, and then propagates freely with a velocity
close to the fast magnetosonic speed in the quiet solar corona.
\end{abstract}



\section{Introduction}

Large-scale coronal disturbances propagating over significant fractions
of the solar disk were for the first time imaged by the Extreme-ultraviolet Imaging
Telescope (EIT) onboard the Solar and Heliospheric Observatory
\citep[SOHO;][]{moses97,thompson98}. After their discovery, these so-called ``EIT waves''
were frequently observed at extreme-ultraviolet (EUV) wavelengths in association
with flares and coronal mass ejections \citep[CMEs; e.g.][]{wills-davey99,klassen00,biesecker02}.
EIT waves were originally assumed to be the coronal counterparts of the
chromospheric Moreton waves \citep{moreton60}, predicted in
Uchida's ``sweeping-skirt-hypothesis'' \citep[][]{thompson99}.
Moreton waves appear as arc-like disturbances in H$\alpha$
filtergrams propagating  away from the flare site with velocities
of 500 to 1000~km~s$^{-1}$ \citep[e.g.][]{warmuth04a}.
They were interpreted as the ground-track signature of a coronal fast-mode
magnetohydrodynamic (MHD) shock wave, which compresses and pushes
the chromospheric plasma downward when sweeping over
it \citep{uchida68}.

However, soon after the discovery of EIT waves, their
interpretation as coronal Moreton waves was questioned, mainly
based on statistical studies reporting significant differences
between both phenomena: in particular EIT waves occur much more
frequently and are on average considerably slower than Moreton
waves with velocities in the range $\sim$170--350~km\,s$^{-1}$
\citep{klassen00,biesecker02}. On the other hand, in case studies
of strong flare/CME events, where both phenomena could be observed
simultaneously, the (single) EIT wave fronts were found to be
roughly co-spatial with the Moreton wave fronts
\citep[e.g.][]{thompson00,warmuth01}. Twelve years after their
discovery, EIT waves (or more generally, global EUV waves) are
still a very controversial subject in solar physics.
Are they really the coronal counterparts of Moreton waves
\citep[e.g.][]{thompson00,warmuth01,vrsnak06,veronig06}. Are they
really (MHD shock) waves
\citep[e.g.][]{wang00,wu01,warmuth01,vrsnak06,wills-davey07}, or
rather propagating disturbances (``pseudo-waves'') related to the
successive opening of magnetic field lines during the CME lift-off
or forced magnetic reconnection ahead of the CME
\citep[e.g.][]{delannee99,chen02,attrill07}? Are they initiated by
the erupting CME, or by the explosive flare energy release?
For a recent review the reader is referred to
\cite{vrsnakANDcliver08}.

A significant drawback in understanding global coronal waves are the
limitations of the EIT observations. Due to the EIT cadence which is
at best 12~min, these events are drastically undersampled. The Extreme
Ultraviolet Imager \citep[EUVI;][]{howard08} instruments onboard the
twin spacecraft of the Solar-Terrestrial Relations Observatory
\citep[STEREO;][]{kaiser08}, provide fundamentally new potential in
studying large-scale disturbances in the solar corona, due to their high
observing cadence in different EUV passbands, the large field-of-view
(up to 1.7~R$_\odot$) and the two separated vantage points. With such
observing capabilities it is for the first time possible to study in
detail the structure and kinematics of global EUV waves together with
the evolution of the associated CME and flare. So far, a few case studies
have been carried out with STEREO/EUVI data. The wave of 19 May 2007
observed at low spacecraft separation was studied in quite some detail
by several groups:
\cite{long08} found that the disturbance was observed in all four EUVI
passbands hinting at the multithermal coronal nature of the
wave. The wave signature in the 304~{\AA} passband was proposed
to be due to a strong contribution from a Si~{\sc xi} line formed at coronal temperatures.
\cite{veronig08} derived a significant deceleration in the
wave kinematics, from about 450 to 200~km~s$^{-1}$, consistent
with the decay of a freely propagating large-amplitude MHD
fast-mode wave \citep[see also][]{long08}.
Further evidence for the fast-mode nature of the
observed disturbance was provided by \cite{gopal09} who studied
the wave reflection at the borders of a coronal hole. Refractions
and reflections of fast-mode coronal MHD waves at regions of high
Alfv\'en velocity were previously studied in simulations
\citep[e.g.][]{wang00,wu01,ofman02}, and provide strong evidence
for the wave nature of the phenomenon. \cite{Patsourakos09a}
presented the first study of an EUV wave observed at large
separation of the two STEREO spacecraft ($\sim$45{\fdg}). These
authors studied the 3D evolution of the associated CME, finding a
close association between the expanding CME loops and the wave
onset. Combining the observations with simulations, these authors
provide strong evidence for the fast-mode MHD wave nature of the
observed disturbance.

In this Letter, we present the first observations of a global EUV
wave observed with the two STEREO spacecraft in quadrature. The
wave of 13 February 2009 was observed simultaneously at disk
center by STEREO-B and on the solar limb by STEREO-A. These
unprecedented STEREO quadrature observations allow us to study in
detail the 3D kinematics and 3D structure of the wave together
with the onset and early evolution of the associated CME, and how
both phenomena are related.

\section{Data}

The STEREO mission consists of two spacecraft with identical
instrument suites orbiting the Sun near the ecliptic plane.
STEREO-A(head) moves on a slightly smaller orbit than the Earth,
while the orbit of STEREO-B(ehind) is slightly larger, resulting
in an increase of their angular separation by 45{\fdg} per year.
On February 13, 2009 the two STEREO satellites were 91{\fdg}
apart, observing in perfect quadrature a global EUV wave
associated with a GOES B2.4 flare/CME event. The evolution of the
coronal wave and the onset of the associated CME is studied in
high-cadence STEREO EUVI imagery. In the 195\,\AA\ filter, the
observing cadence was 10~min for both STEREO satellites, in the
171\,\AA\ filter it was 2.5~min for STEREO-A and 5~min for
STEREO-B [henceforth ST-A and ST-B]. To compare the observations
from ST-A (on 13 February 2009, at a distance of 0.964\,AU from
Sun) and ST-B (1.003\,AU), all images were re-scaled to Earth
distance (0.987\,AU). The EUVI data were fully reduced
using the SECCHI$\underline{~}$PREP routines available within Solarsoft,
and differentially rotated to 05:30~UT. To enhance the contrast of the
transient wave signatures, we constructed from the EUVI 171 and 195~{\AA} frames
running difference and running ratio images, where we subtracted (divided)
from each image a frame that was taken 10~min before.



\section{Results}

\subsection{Wave kinematics} \label{kinematics}

\begin{figure}[p]
\resizebox{16cm}{!}{\includegraphics[angle=0]{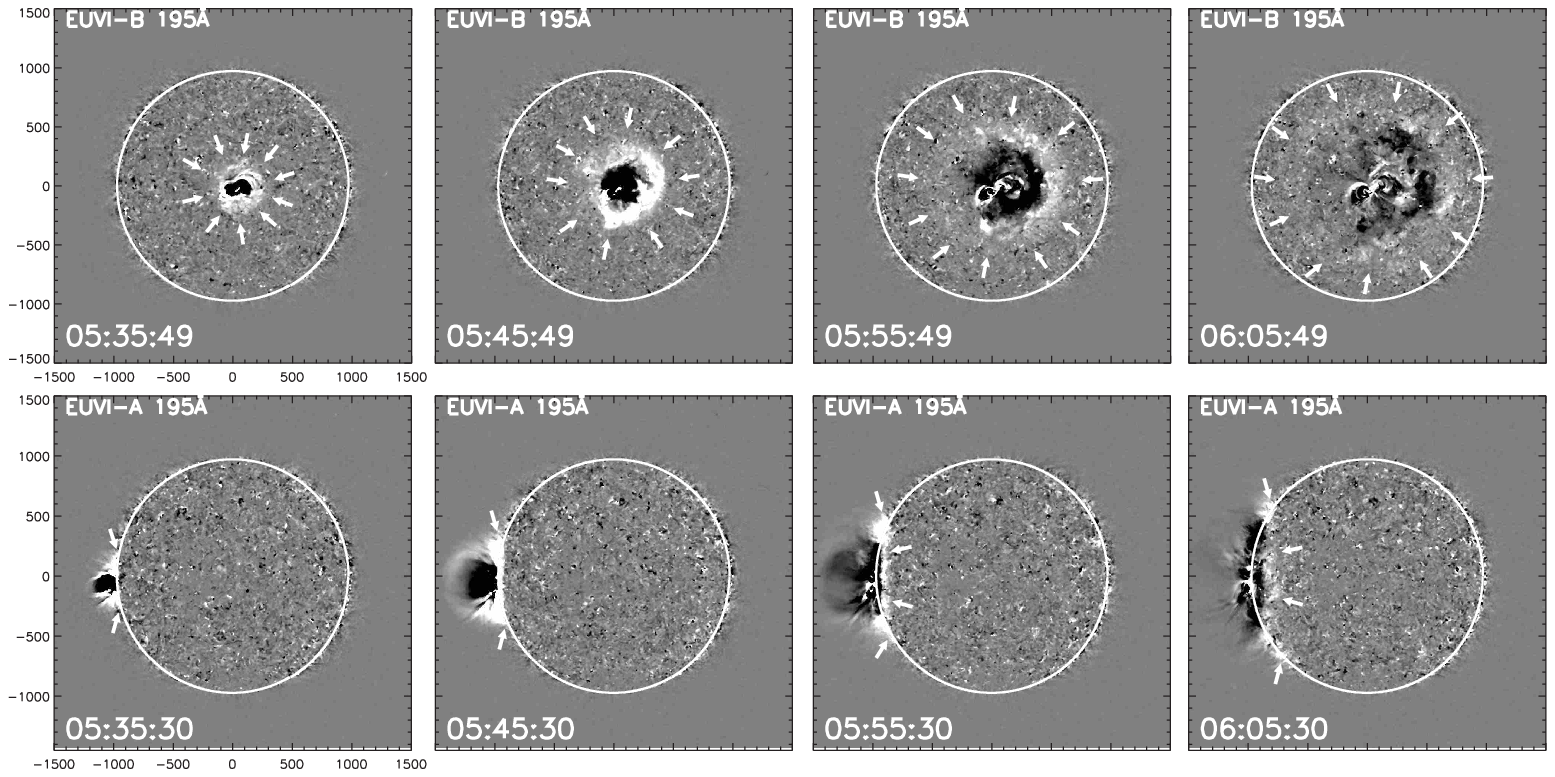}}
\caption{Sequence of median-filtered running difference images
recorded in the EUVI 195\,\AA\ channel with a cadence of 10~min.
The coronal wave (outlined by arrows) is observed on-disk in ST-B
(top) and on the limb in ST-A (bottom). Axes units are in arcsec.
[See also the accompanying
movie.]} \label{fig1a}
\end{figure}

\begin{figure}[p]
\includegraphics{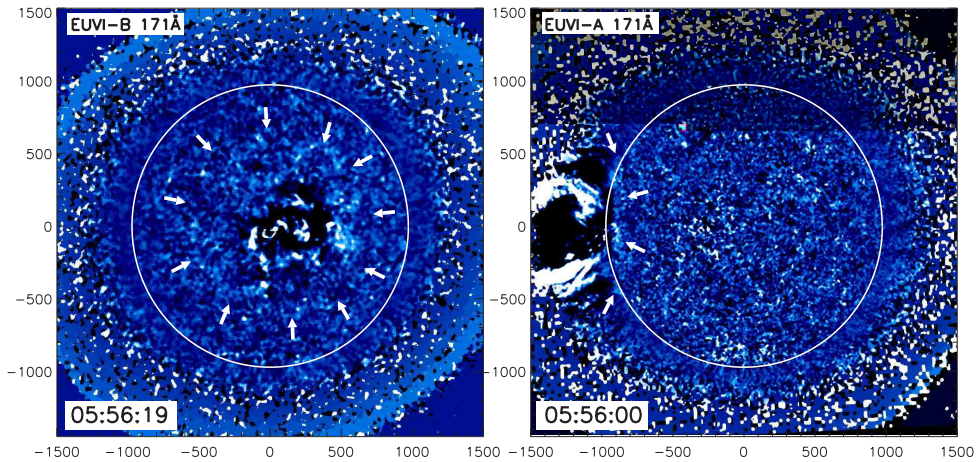}
\caption{Median-filtered running ratio images taken in the EUVI 171\,\AA\ channel by ST-B (left) and ST-A
(right) at 05:56~UT. Axes units are in arcsec. The wave evolution can be followed in the
accompanying movie. }\label{fig2a}
\end{figure}

ST-B observed the coronal wave on the solar disk, with the wave
center and the associated weak B2.4 flare being located precisely
at Sun center. Due to the satellite's quadrature configuration,
this results in a perfect limb event for ST-A. Figure \ref{fig1a}
shows the wave evolution in simultaneous EUVI 195\,\AA\ running
difference images from ST-A and ST-B (see also the accompanying
movie). Figure~\ref{fig2a} shows one snapshot of the wave observed
in the EUVI 171\,\AA\ filter. A movie of 171\,\AA\ running ratio
images (with 5~min cadence) is available in the online material.
The coronal wave was first observed at 05:35\,UT, and could be
followed up to at least 06:20\,UT. ST-B observations reveal that
the coronal wave is globally propagating into all directions, but
with much larger brightness toward the western
hemisphere. ST-A shows the wave propagation at the solar limb as
well as the erupting CME structure. In the last two panels in Fig.~\ref{fig1a}, ST-A observes the coronal
wave simultaneously above the limb as well as on the disk.

\begin{center}

\begin{figure}
\epsscale{0.9} \plotone{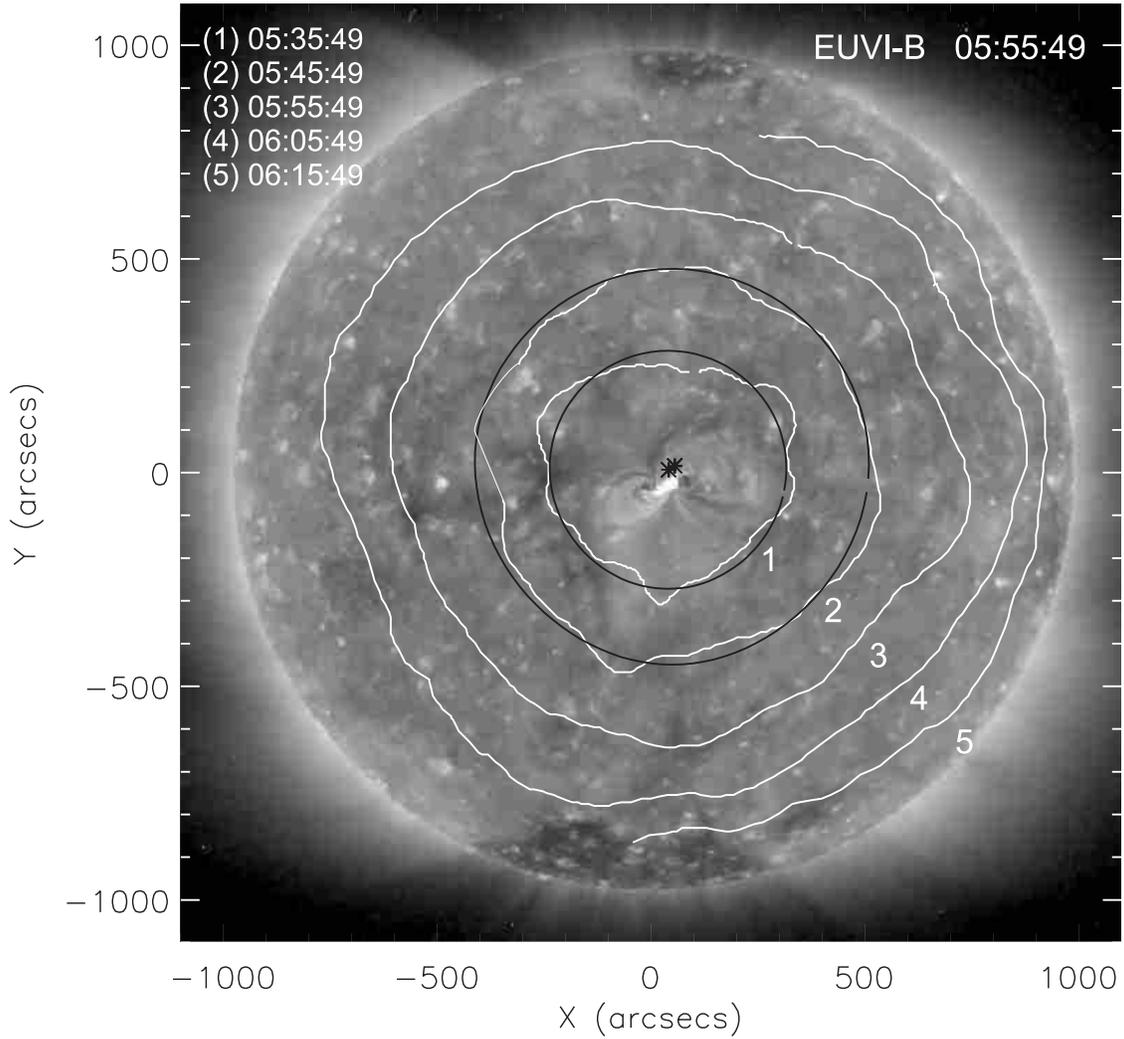}
\caption{EUVI-B 195\,\AA\ image recorded at 05:55:49~UT together with the wavefronts derived from 195\ \AA\ difference images during 05:35:49 to 06:15:49~UT (white lines).
Dark solid lines indicate circular fits to the earliest observed wavefronts, from which the wave front center was derived.
Note that the centers derived from the different wavefronts (indicated by black crosses) lie close to each other near
the northern edge of the small flare loop arcade.
\label{fig3a}}
\end{figure}
\end{center}

In Fig.~\ref{fig3a}, we plot a ST-B 195\,\AA\ direct image
together with all wavefronts determined from ST-B 195\,\AA\
running difference images. The source center required to derive
the kinematics of the wave was obtained from circular fits to the
earliest wavefronts in the 3D spherical plane \citep[for details
we refer to][]{veronig06}. The resulting fits and derived wave
centers are also plotted in Fig.~\ref{fig3a}, showing the wave
initiation center slightly north of the flaring active region. For
each wavefront identified in the ST-B 195\,\AA\ and 171\,\AA\ running difference images,
we determined the mean distance from the
derived center along the spherical solar surface.

The bottom panel in Fig.~\ref{fig_kin} shows the resulting wave
kinematics together with the GOES soft X-ray flux of the
associated B2.4 flare. Green triangles and blue diamonds mark the
mean distances of the wavefronts observed on-disk in EUVI-B
195\,\AA\ and in 171\,\AA,\ respectively.
The velocity of the wave, $v \sim 243\pm 14$~km~s$^{-1}$, is constant over the full propagation range from 200 to 750 Mm.
We also studied the wave kinematics separately for different
propagation sectors (not shown in Fig.~\ref{fig_kin}). The
velocity derived for different directions is the same within the
error bars but the brightness is quite different for the
propagation into the western and eastern direction (see
Figs.~\ref{fig1a} and the online movies). These findings imply
that the propagation speed does not vary with the
amplitude of the perturbation.

\subsection{3D structure of the wave} \label{3D}

The lateral observations of the wave from ST-A clearly reveal its
3D nature, with the wave front visible to a considerable
height above the solar surface (last panel in Fig.~\ref{fig5a}).
In the following, we compare the wave kinematics as observed
on-disk from ST-B with those derived from the limb-observations in
ST-A. Since ST-A has an edge-on view of the wave, the distance
measurements are basically free from projection effects. We also
show how the data from two observing platforms separated by
$\sim$90$\fdg$ can be used to estimate the typical heights of the
wave observations in on-disk measurements.

The red asterisks in the bottom panel in Fig.~\ref{fig_kin} show
the wave kinematics as obtained from ST-A by measuring the wave
evolution along the solar limb, i.e.\ at a height of 0$''$. The
resulting data points lie 50--100~Mm behind the kinematical curve
derived from the on-disk ST-B observations, indicating that the
wave propagates significantly above the solar surface. However, if
we just increase the spherical surface on which the wavefront
observed by ST-A is measured, we are not able to obtain a
reasonable agreement with the wave kinematics derived from ST-B
on-disk observations. This is demonstrated by the red x~symbols in
Fig.~\ref{fig_kin}, showing the wave kinematics derived along a
spherical surface at a height of $130''$ above the solar limb.
Considering the 3D nature of the wave, projection effects have to
be taken into account in the ST-B on-disk observations. The ST-B
on-disk observations are then a result of integrating the emission
from the wave front high above the solar surface along the
line-of-sight (illustrated by the long arrow in Fig.~\ref{fig5a}).
Assuming a height of 130$''$ ($\sim$90~Mm) for the wave
observations projected onto the solar surface, the wavefront
distances and velocities derived from ST-A and ST-B are in
agreement (see Fig.~\ref{fig_kin}).

What do we learn from this? First, the emission of EUV waves
observed on-disk originates from high above the solar surface
($\sim$80--100~Mm). This is similar to the height of
$\sim$$90\pm 7$~Mm derived for a global EUV wave observed with
STEREO at a spacecraft separation of 45{\fdg} using triangulation
techniques \citep{Patsourakos09a}. Such heights are comparable to
the coronal scale-height of 50--100~Mm for quiet Sun temperatures
of 1--2~MK, and consistent with the propagation of a MHD fast-mode
wave over quiet Sun regions, since the wave perturbs and
compresses the ambient coronal plasma with its bulk confined
within a coronal scale-height \citep{Patsourakos09a}. Second, the
derived wave height of about 80--100~Mm is not the upper edge (see
also the right panel in Fig.~\ref{fig5a}) but reflects the maximum
of the line-of-sight integration of the increased intensity due to
the plasma compression at the wave front. Third, there may be an
underestimation of about 10\% of the coronal wave speed derived
from the disk-observations ($243\pm 14$~km~s$^{-1}$) with regard
to the ``real'' propagation velocity at heights of 80--100~Mm,
where we find $v \sim 263\pm 16$~km~s$^{-1}$.

\subsection{Associated coronal mass ejection} \label{driver}

The early evolution of the erupting CME associated with the
coronal wave is observed with high-cadence by ST-A, basically free
from projection effects. This enables us to study the expansion of
the erupting CME with relation to the wave propagation. In order
to enhance the coronal structures above the limb, we removed the
steep radial intensity gradient by a normalizing-radial-graded
filter \citep[NRGF;][]{morgan06}. A bulb-like pre-CME structure is
observed hours before the event, and starts to erupt around
05:30--05:35\,UT (Fig.~\ref{fig5a}).

In order to follow in detail the CME take-off and lateral
expansion in the low corona, we selected various vertical slices
in the plane of sky from the NRGF-filtered EUVI-A 171\,\AA\ images
(available at a cadence of 2.5~min) at different coronal heights.
In Fig.~\ref{fig_kin}, we show three stack plots extracted from
$2''$-slices centered at $x=-1000''$, $-1100''$, and $-1200''$
(indicated in the first panel in  Fig.~\ref{fig5a}) covering the
length $y=[-600'', +500'']$. These plots reveal the pre-existing
CME structure as well as the expansion of the CME flanks.
(Though we note that the stack plots include also some contribution
of the CME upward movement in addition to the expansion.)
The bifurcation of the bright horizontal feature indicates the onset
of the lateral expansion of the CME. In the $x=-1100''$ ($H \sim
90$~Mm) stack plot, the bifurcation starts at $\sim$05:32\,UT with
the fastest growth around 05:35\,UT. For $x=-1200''$ ($H \sim
160$~Mm), the fast onset occurs later by about 2--3~min. The
expansion stops at $\sim$05:50~UT for $x=-1100''$ and at
$\sim$06:00~UT for $x=-1200''$. In the $x=-1000''$ slice, which is
low in the corona ($H \sim 20$~Mm), only a marginal CME
expansion is observed, revealing the anchored ends of the CME
flanks.

Studying stack plots at various heights ($x$-positions), we find
that the coronal wave onset ($\sim$05:35\,UT) is consistent with
the CME lateral expansion at coronal heights in the range
70--120~Mm. At the slice $x=-1100''$ ($H \sim 90$~Mm), shown in
Fig.~\ref{fig_kin}, the expansion starts around 05:32 reaching the
fastest growth at $\sim$05:35\,UT. This suggests that the wave is
initiated between $\sim$05:32--05:35\,UT by the CME expanding
flanks, and thereafter propagates freely. In the last panel in
Fig.~\ref{fig5a}, we show a EUVI-A 195\,\AA\ high-contrast ratio
image at 06:05\,UT, which clearly reveals the coronal wave (above
the limb as well as on the disk) far away from the erupting
structure.

\section{Discussion and Conclusions}

From the STEREO quadrature observations of a coronal wave
observed simultaneously in frontal and lateral view, we find that
the EUV emission of the coronal wave originates at a height of
about 80--100~Mm. The wave propagation speed (corrected for
projection effects) is $\sim$260~km~s$^{-1}$. The wave speed is
constant over the full range of quiet-Sun region it traversed and
homogeneous in all directions though the perturbation amplitudes
are quite different for different propagation directions (in
particular toward the western and eastern direction as observed
from ST-B). This is a different behavior compared to the wave
observed by STEREO on 19 May 2009, which revealed a distinct
deceleration from $\sim$450 to 200~km~s$^{-1}$ \citep{veronig08}.
The associated flare X-ray flux is very weak (GOES B2.3), which
suggests that the flare pressure pulse is unlikely to be causing
the wave.

\begin{figure}
\epsscale{0.58} \plotone{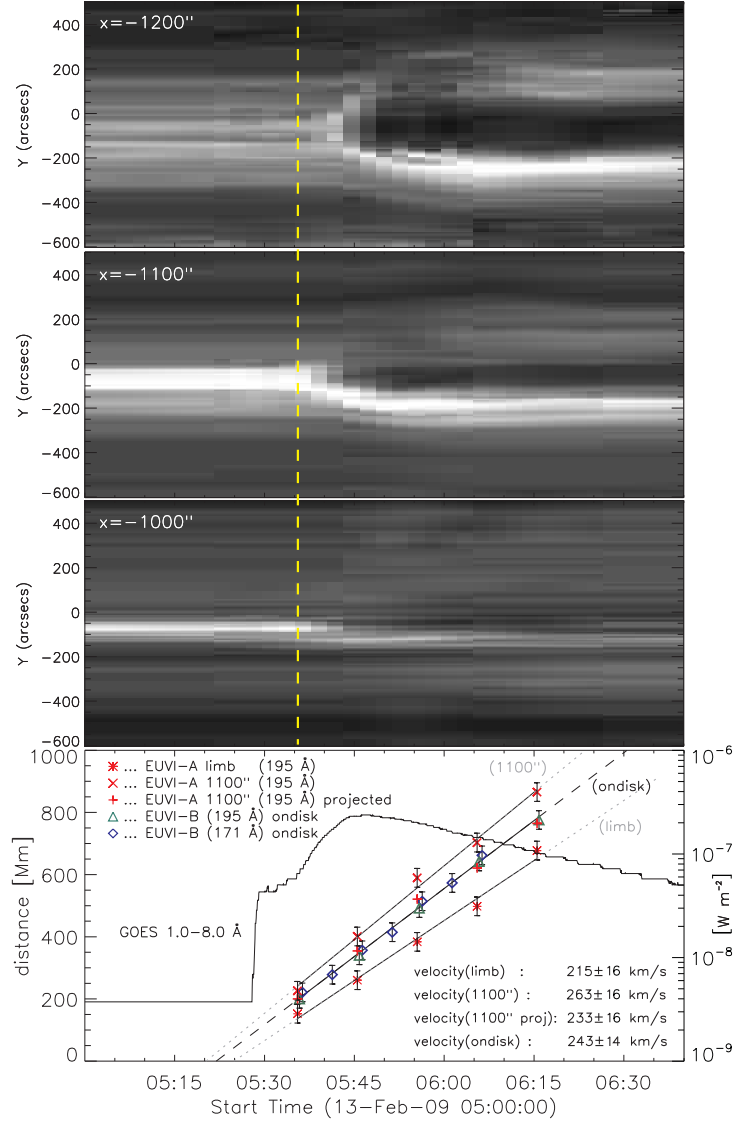}
\caption{Top panels: Stack plots derived from EUVI-B 171\,\AA\ images (2.5~min cadence).
The three panels show cuts taken at different heights across
the erupting CME structure (cf.\ Fig.~\ref{fig5a}). The vertical line indicates the time of the first observation of the
coronal wave.
Bottom panel: Kinematics of the coronal wave studied on-disk in EUVI-B and on the limb in
EUVI-A together with linear fits and the GOES 1--8\,\AA\ flux of the associated flare. Green triangles and blue diamonds mark the wave kinematics observed on-disk by EUVI-B. Red symbols indicate the wave kinematics derived from EUVI-A: red asterisks by measuring the propagated distance along the spherical solar surface (i.e.\ along the limb), red x~symbols when measuring the distance of the wavefront along the circle with radius 1100$''$ (i.e.\ at a height of 130$''$ above the solar limb), and red crosses when measuring the distance of the wavefront at a height of 130$''$ projected onto the solar surface (as illustrated on the right panel of Fig.~\ref{fig5a}).\label{fig_kin}}
\end{figure}

\begin{figure*}
\epsscale{1} \plotone{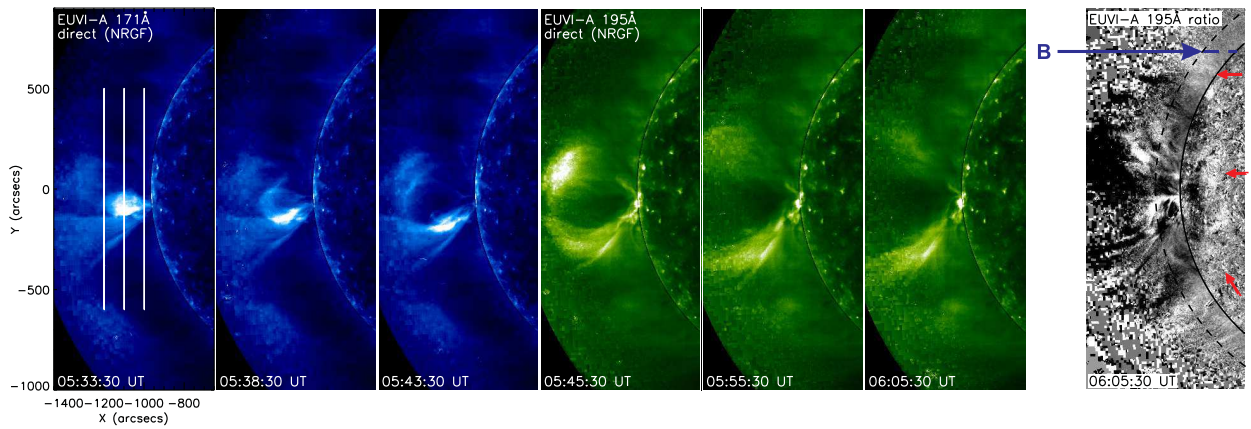} \caption{NRGF-filtered EUVI-A 171~\AA\ and 195~\AA\ images showing the early evolution of the erupting CME. The vertical lines in the first image indicate the positions at which the stack plots shown in Fig.~\ref{fig_kin} were derived. The last panel shows a high-contrast ratio image between two consecutive EUVI-A 195~\AA\ frames (06:05~UT/05:55~UT) revealing the coronal wave as well as the erupting CME structure. The solar limb (black circle) and the estimated observation height of the wave above the solar surface (dashed circle with radius of 1100$''$, i.e. $\sim$130$''$ above the solar limb) are indicated. The on-disk signature of the coronal wave observed by EUVI-A is pointed out by the red arrows. The projected view from STEREO-B is illustrated by a blue arrow. \label{fig5a}}
\end{figure*}

The first observations of the wave coincides with the onset of
fast expansion of the CME expanding flanks at a height of
$\sim$$70$--$120$~Mm (see Fig.~\ref{fig_kin}) suggesting that this
expansion initiates the wave. Afterwards, the wave propagates
freely and is far ahead of the observed CME flanks (see right
panel of Fig.~\ref{fig5a}). Such a behavior was also observed in
the EUVI event studied in \cite{Patsourakos09a}. This offset of
the wave front with respect to the projected CME flank is not in
line with the behavior expected from the ``pseudo-wave'' models
\citep[e.g.][]{delannee99,chen02,attrill07}, where the wave front
should be basically co-spatial with the CME front. From the stack
plots in Fig.~\ref{fig_kin} and from direct measurements of the CME expanding
flanks, we find that the peak velocity of the lateral CME expansion
at a fixed height of $\sim$90~Mm is $\sim$150~km~s$^{-1}$.
This provides us with an estimate of the driver speed.

These findings (driver speed smaller than wave speed; constant
wave velocity over the full observation range, i.e.\ not varying
with amplitude) indicate that we are dealing with a fast-mode MHD
wave not too far from the linear regime. In this case, we would
expect the wave to propagate at the characteristic speed of the
medium (i.e.\ the fast magnetosonic speed), independent of the
amplitude of the perturbation. A fast magnetosonic speed in the
range of 260~km~s$^{-1}$ is not unreasonable for quiet Sun coronal
conditions in the deep solar minimum. Recently, direct magnetic
field measurements of the corona were deduced by \cite{lin04} who
obtained a field strength of 4 Gauss above an active region at a
height of $\sim$1.1~R$_\odot$. Assuming for quiet Sun conditions a
magnetic field strength in the range of 1 to 3 Gauss and a coronal
density of $5 \cdot 10^8$~cm$^{-3}$, the fast magnetosonic speed
lies in the range 210 to 350~km~s$^{-1}$ which is consistent with
the observed wave velocity. 
Thus, we conclude that the observed EUV transient is a real wave, and that
its observational characteristics is consistent with a low-amplitude MHD
fast-mode wave (i.e.\ not a shock wave). Finally, we note that there is a recent paper on
the same event \citep{Patsourakos09b}. These authors basically
arrive at the same conclusions of a fast-mode MHD wave but use a
different approach, modeling the CME and wave structures based on
EUVI and COR observations.

\acknowledgments I.W.K. and A.M.V. acknowledge the Austrian Fonds
zur F\"orderung der wissenschaftlichen Forschung (FWF grant
P20876-N16). M.T. is a recipient of an APART-fellowship of the
Austrian Academy of Sciences at the Institute of Physics,
University of Graz (APART 11262). We thank the STEREO/SECCHI teams
for their open data policy and Dr. B. Vr\v{s}nak for insightful
discussions.

\bibliographystyle{apj}


\end{document}